\documentclass[twocolumn,twocolappendix]{aastex631}

\usepackage{natbib}
\usepackage{graphicx}	
\usepackage{amsmath}	
\usepackage{amssymb}	
\usepackage{bm}	
\usepackage{subfigure}
\usepackage{longtable}
\usepackage{array}
\usepackage{booktabs}
\usepackage{multirow}
\usepackage{xcolor}
\usepackage{lipsum}

\newcommand{\msun}{\,{M_{\odot}}}

\shorttitle{The blue KN in GRB~211211A}
\shortauthors{Hamidani, Tanaka, Kimura, Lamb, and Kawaguchi}

\begin{document}
\title{GRB 211211A: The Case for an Engine Powered over r-Process Powered Blue Kilonova}

\correspondingauthor{Hamid Hamidani}
\email{hhamidani@astr.tohoku.ac.jp}

\author[0000-0003-2866-4522]{Hamid Hamidani}
\affiliation{Astronomical Institute, Graduate School of Science, Tohoku University, Sendai 980-8578, Japan}

\author[0000-0001-8253-6850]{Masaomi Tanaka}
\affiliation{Astronomical Institute, Graduate School of Science, Tohoku University, Sendai 980-8578, Japan}

\author[0000-0003-2579-7266]{Shigeo S. Kimura}
\affiliation{Frontier Research Institute for Interdisciplinary Sciences, Tohoku University, Sendai 980-8578, Japan}
\affiliation{Astronomical Institute, Graduate School of Science, Tohoku University, Sendai 980-8578, Japan}

\author[0000-0001-5169-4143]{Gavin P. Lamb}
\affiliation{Astrophysics Research Institute, Liverpool John Moores University, Liverpool Science Park IC2, 146 Brownlow Hill, Liverpool, UK, L3 5RF}

\author[0000-0003-4443-6984]{Kyohei Kawaguchi}
\affiliation{Max Planck Institute for Gravitational Physics (Albert Einstein Institute), Am M\"{u}hlenberg 1, Potsdam-Golm, 14476, Germany}
\affiliation{Center of Gravitational Physics and Quantum Information,
 Yukawa Institute for Theoretical Physics, 
Kyoto University, Kyoto, 606-8502, Japan} 

\begin{abstract}
The recent Gamma-Ray Burst (GRB) GRB~211211A provides the earliest ($\sim 5$ h) data of a kilonova (KN) event, displaying bright ($\sim10^{42}$ erg s$^{-1}$) and blue early emission. 
Previously, this KN has been explained using simplistic multi-component fitting methods. 
Here, in order to understand the physical origin of the KN emission in GRB~211211A, we employ an analytic multi-zone model for r-process powered KN. 
We find that r-process powered KN models alone cannot explain the fast temporal evolution and the spectral energy distribution (SED) of the observed emission. 
Specifically, i) r-process models require high ejecta mass to match early luminosity, which overpredicts late-time emission, while ii) red KN models that reproduce late emission underpredict early luminosity.
We propose an alternative scenario involving early contributions from the GRB central engine via a late low-power jet, consistent with plateau emission in short GRBs and GeV emission detected by Fermi-LAT at $\sim10^4$ s after GRB 211211A. 
Such late central engine activity, with an energy budget of $\sim \text{a few }\%$ of that of the prompt jet, combined with a single red-KN ejecta component, can naturally explain the light curve and SED of the observed emission; 
with the late-jet -- ejecta interaction reproducing the early blue emission and r-process heating reproducing the late red emission. 
This supports claims that late low-power engine activity after prompt emission may be common. 
We encourage early follow-up observations of future nearby GRBs, and compact binary merger events,
to reveal more about the central engine of GRBs and r-process events.
\end{abstract}


\section{Introduction}
\label{sec:1}
Traditionally, gamma-ray burst (GRB) are classified into two classes based on their duration ($T_{\rm{90}}$): Long (LGRB; $T_{\rm{90}}> 2$ s) and Short (SGRB; $T_{\rm{90}}< 2$ s) (\citealt{1993ApJ...413L.101K}).
On one hand, LGRBs are explained by the collapse of massive stars (collapsar model; \citealt{1999ApJ...524..262M});
and in fact, most nearby LGRBs (with a few exceptions; GRB~060614; GRB~060505; etc.) are associated with bright supernova (SN) explosions confirming 
this scenario (\citealt{1998Natur.395..672I,2003ApJ...591L..17S}).
On the other hand, SGRBs have theoretically been associated with binary neutron star (BNS; also black hole-neutron star BH-NS) mergers (\citealt{1986ApJ...308L..43P}; \citealt{1986ApJ...308L..47G}).
This scenario is consistent with observations that show spatial off-sets between SGRB locations and their candidate host galaxies (\citealt{2010ApJ...708....9F,2022ApJ...940...56F,2022ApJ...940...57N,2022MNRAS.515.4890O}).

Moreover, BNS mergers are a site of r-process nucleosynthesis whose radioactivity powers an optical-infrared transient referred to as ``kilonova/Macronova" (KN hereafter) (\citealt{1998ApJ...507L..59L}; \citealt{2005astro.ph.10256K}; \citealt{2010MNRAS.406.2650M}).
This was confirmed with the gravitational wave (GW) and electromagnetic (EM) observations of GW170817, associating a BNS merger event with a SGRB (GRB~170817A) and with the KN transient AT2017gfo (e.g., \citealt{2017PhRvL.119p1101A}; \citealt{2017ApJ...848L..13A}; \citealt{2017Sci...358.1570D}; \citealt{2017Sci...358.1559K}; \citealt{2017PASJ...69..102T}).

Observations of GRB~211211A show a long main peak ($\sim13$ s; according to Fermi-GBM), followed by a softer and smoother extended emission ($\sim 55$ s long; Fermi-GBM) (\citealt{2022Natur.612..223R,2022Natur.612..232Y,2022Natur.612..228T,2023ApJ...954L...5V}).
A candidate host galaxy was identified, allowing for redshift ($z =0.0762$) and distance ($d_{\rm{L}} = 346$ Mpc) measurements (\citealt{2022Natur.612..223R,2022Natur.612..228T}).
As a bright nearby GRB, GRB~211211A was the target of many follow-up observations.
However, although according to the traditional classification scheme GRB~211211A is a LGRB, no sign of a supernova (SN) could be found, while a clear KN transient was identified (\citealt{2022Natur.612..223R,2022Natur.612..228T}).
This suggests that GRB~211211A originated from a BNS/BH-NS merger event, and that SGRBs' engine activity can last longer than the nominal 2 s duration limit (see Figure 2 in \citealt{2015ApJ...804L..16K}; \citealt{2022ApJ...934L..12G,2023ApJ...958L..33G}; also see \citealt{2023arXiv230800633G,2024Natur.626..742Y,2024Natur.626..737L} for the similar event GRB~230307A).

Thanks to its nearby location, the KN that followed GRB~211211A was observed at times earlier than any event before (\citealt{2022Natur.612..223R,2022Natur.612..228T}).
These observations revealed a bright early blue KN 
($\sim 3\times 10^{42}$ erg s$^{-1}$ at $\sim 5$ h; \citealt{2022Natur.612..228T}).
The origin of this blue KN (also in the GW170817/AT2017gfo event; \citealt{2017Sci...358.1559K,2017Sci...358.1570D}) is not well understood, considering that KNe are expected to peak in optical-infrared bands as they contain substantial fractions of heavy elements (i.e., lanthenides, with high opacities).
Through the fitting of photometric data, it has been shown that the KN is well-explained by two or three ejecta components (red, blue, and purple) with given masses, opacities, and velocities \citep{2022Natur.612..223R}.
However, these methods (e.g., \citealt{2017ApJ...851L..21V})
are based on 
one-zone models (\citealt{1982ApJ...253..785A}).
Additionally, parameter fitting results are often at odds with first-principle numerical relativity simulations (see \citealt{2017PhRvD..96l3012S,2018ApJ...865L..21K,2018ApJ...860...64F,2019EPJA...55..203S}).
Therefore, it is crucial to investigate the origin of the blue KN emission using more 
sophisticated models.

Another key observation related to GRB~211211A is the detection of high-energy ($\sim 0.1-1$ GeV) photon emission by Fermi-LAT $\sim 10^4$ s after the prompt emission (\citealt{2022Natur.612..236M,2022ApJ...933L..22Z}). 
This is the first time that late time GeV emission from a supposed BNS merger event has been detected with high significance at $>5\sigma$ (GRB~160821B is another similar event where sub-TeV emission was detected by MAGIC, although less significantly at $\sim 3\sigma$;  \citealt{2019ApJ...883...48L,2021ApJ...908...90A,2021ApJ...908L..36Z}).
The GeV emission was explained by KN photons interacting (via inverse Compton scattering) with low-power jet powered by the central engine long after the prompt phase (\citealt{2022Natur.612..236M}). \cite{2022ApJ...933L..22Z} suggested that the afterglow model could explain this emission, however, this requires extreme jet parameters ($E_{\rm{k,iso}}\sim 10^{53}$ erg and $\theta_j\sim 1^\circ$).

The idea that the GRB engine stays active long after the prompt emission is not new\footnote{There is a similar argument for LGRBs based on observations of X-ray flares (\citealt{Burrows2005Sci...309.1833B,Nousek2006ApJ...642..389N}).}; 
observations have consistently shown that $\sim 10^2 - 10^4$ s after the prompt emission, bright X-ray emission that cannot be explained by the standard afterglow model is emitted (see \citealt{2005Natur.438..994B}; \citealt{2013MNRAS.431.1745G}; \citealt{2017ApJ...846..142K}; \citealt{2019ApJ...877..147K}; etc.).
These late phases are referred to as ``extended" (with $L_{\rm{X}}\sim 10^{48}$ erg s$^{-1}$ for $\sim 10^2$ s) or ``plateau" (with $L_{\rm{X}}\sim 10^{46}$ erg s$^{-1}$ for $\sim 10^4$ s) phases, 
and are typically associated with late engine activity (\citealt{2005ApJ...631..429I}; \citealt{2015ApJ...804L..16K}; \citealt{2023ApJ...958L..33G}; etc.).
These late phases of engine activity might be ubiquitous in SGRBs (\citealt{2017ApJ...846..142K})\footnote{For GW170817/GRB~170817A, these late phases could not be confirmed due to the off-axis line-of-sight and the earliest follow-up observations being too late (\citealt{2017Sci...358.1565E}).}.

In an effort to understand the origin of KN transients, we revisit GRB~211211A, which provides the earliest data of a KN to date. 
We investigate the source of the bright early blue KN emission via analytic modeling, and test the hypothesis that this blue emission is r-process powered.
Additionally, we explore the impact of a late low-power (i.e., plateau) jet interacting with the merger ejecta has on the KN emission.

This paper is organized as follows. 
In Section \ref{sec:3} we present our physical model for r-process powered KN.
In Section \ref{sec:4} we present our results and explain the limitations of the r-process powered KN scenario.
An alternative scenario of central engine powered KN is presented in Section \ref{sec:5}.
Finally, a discussion and conclusion are presented in Section \ref{sec:6}.
Details related to GRB~211211A's data can be found in
Appendix \ref{data}.


\section{Method}
\label{sec:3}
\subsection{R-process powered kilonova model}
\label{sec:KN model}
We consider the same KN model as in \cite{2024ApJ...963..137H} (see their Appendix E), with additional improvements.
The main approximations of the model are:
\begin{itemize}
    \item The ejecta is expanding homologously with $\beta_{\rm{m}}$ and $\beta_{\rm{0}}$ as the maximum and minimum velocities (in units of $c$), respectively.
    \item The density profile of the ejecta is approximated to single power-law with $n$ as its power-law index: 
        \begin{equation}
        \rho\propto \beta^{-n} .
    \end{equation}
    This is significantly more realistic than the widely used KN models (e.g., see Section 3 in \citealt{2017ApJ...851L..21V}), as the time evolution of optical depth, photospheric radius, luminosity, and temperature, depend on the spatial distribution of density.
    \item The time evolution of r-process energy deposition per mass ($\dot{\varepsilon}$) is approximated to a power-law function: 
    \begin{equation}
        \dot{\varepsilon} = \dot{\varepsilon}_{\rm{0}}\left(\frac{t}{\rm{1\:d}}\right)^{-k} ,
    \end{equation}
    with $k=1.3$, and $\dot{\varepsilon}_{\rm{0}}= 2\times 10^{10}$ erg g$^{-1}$ s$^{-1}$ (i.e.,  $\dot{\varepsilon}_{\rm{0}}\sim 1$ MeV nuc$^{-1}$ s$^{-1}$ at $t=0.1$ s; \citealt{2014ApJ...789L..39W,2021ApJ...922..185I}).
    \item At early times (first a few days) r-process energy deposition is assumed to be dominated by beta-decay (\citealt{2014ApJ...789L..39W,2019ApJ...876..128K})\footnote{This is a conservative consideration as it gives a faster time evolution for $f_{\rm{tot}}$ (without alpha and fission contributions), and considering our aim of exploring alternative scenarios that explain the fast time evolution in the light curve of the KN associated with GRB~211211A.}. 
    Hence, we adopt an analytic time dependent thermalization efficiency term $f_{\rm{tot}}(t)$ (\citealt{2016MNRAS.459...35H,2016ApJ...829..110B}) following the analytic model in \cite{2019ApJ...876..128K} [see their equation (51)]. 
    In reality $f_{\rm{tot}}(t)$ should also have spatial dependency, but here for simplicity we consider a one-zone prescription.
    \item A sharp diffusion shell at:  
    \begin{equation}
        \tau =c/(v_{\rm{m}} -v_{\rm{d}}) ,
        \label{eq:tau}
    \end{equation}
    is adopted, where $\beta_{\rm{m}}=v_{\rm{m}}/c$ and $\beta_{\rm{d}}=v_{\rm{d}}/c$ are the outer velocity of the ejecta and the velocity of the sharp diffusion shell, respectively (\citealt{2012ApJ...747...88N,2015ApJ...802..119K,2023MNRAS.524.4841H}).
    \item Grey opacity is adopted:
\begin{equation}
        \kappa = \rm{Const.}
\end{equation}
    with $\kappa$ values taken from realistic radiative transfer simulation results (see \citealt{2023arXiv230405810B}).
    \item In the first a few days the KN emission is approximated to a blackbody (\citealt{2015ApJ...802..119K,2018MNRAS.481.3423W}).
\end{itemize}

For a given shell in the ejecta with a velocity $\beta$, density, optical depth, and thermal energy density can be found as a function of time.
Then, with the photon diffusion criteria, emission can be found analytically as a function of time.
Therefore, for a given set of the following parameters KN emission (light curve and spectral energy distribution; SED) can be found analytically: ejecta mass ($M_{\rm{e}}$), power-law index of the density profile of the ejecta ($n$), maximum ejecta velocity ($\beta_{\rm{m}}$), minimum ejecta velocity ($\beta_{\rm{0}}$), and grey opacity ($\kappa$).

There are two distinct terms that contribute to the KN emission, as follows [equation (E3) in \citealt{2024ApJ...963..137H}]:
\begin{equation}
    L_{\rm{KN}}(t)=L_{\rm{KN}}(<\beta_{\rm{d}},t)+L_{\rm{KN}}(\geqslant\beta_{\rm{d}},t) ,
        \label{eq:Lbl KN}
\end{equation}
where $t$ is time ($t\approx t_{\rm{obs}}$), and $\beta_{\rm{d}}$ is the velocity of the diffusion shell.
The first term [$L_{\rm{KN}}(<\beta_{\rm{d}},t)$] is emission due to leaking of trapped thermal (or internal) energy as the diffusion shell moves inward (in a Lagrangian coordinate) through the optically thick part of the ejecta; we refer to it as the ``diffusion" term.
The second term [$L_{\rm{KN}}(\geqslant\beta_{\rm{d}},t)$] is emission due to instantaneous deposition of thermal energy (via r-process heating) in the optically thin part of the ejecta; we refer to it as the ``deposition" term. 

\begin{figure}
    \centering
    \includegraphics[width=0.99\linewidth]{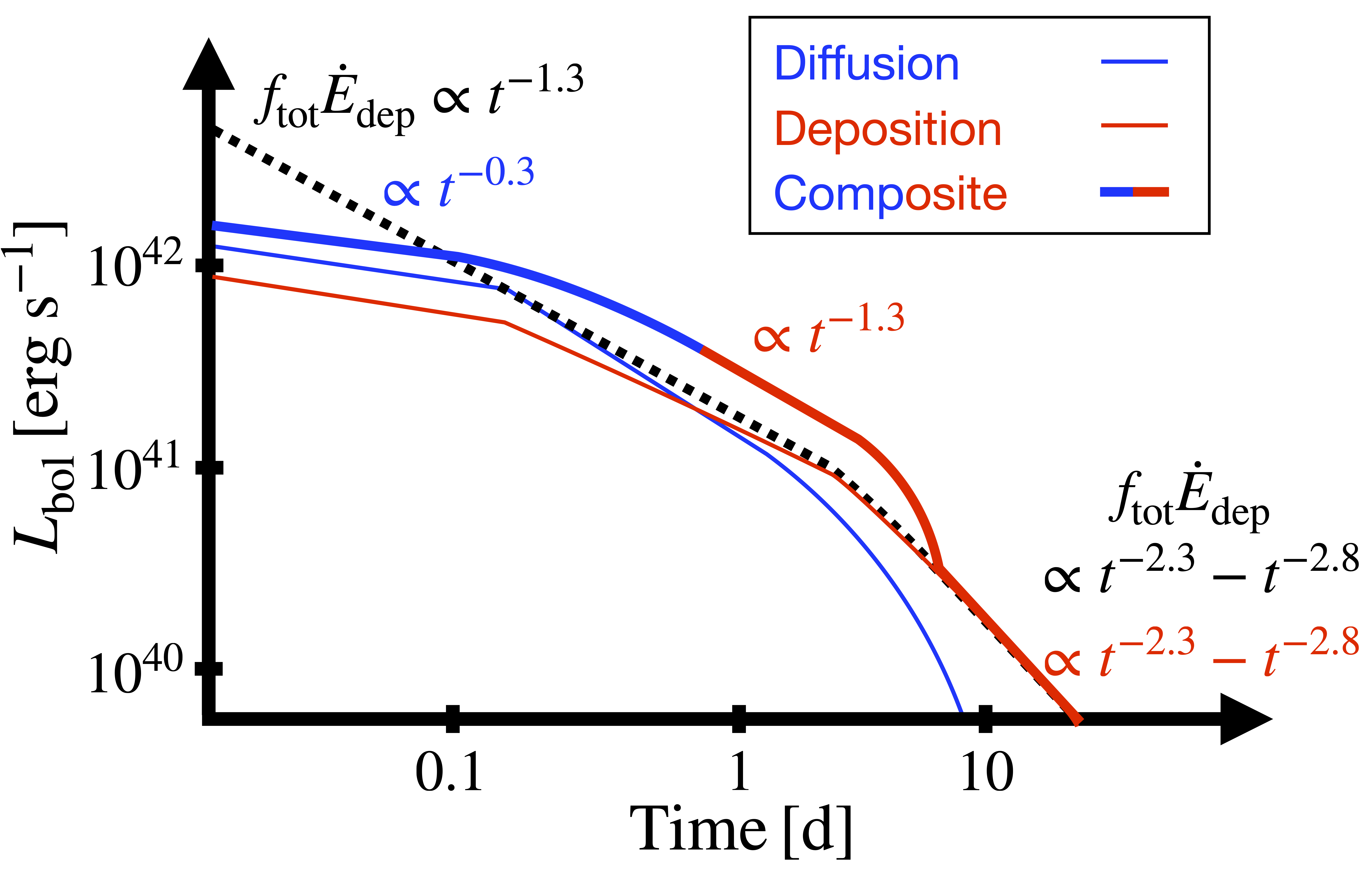} 
  \caption{Time evolution of r-process powered KN light curve. The early blue KN emission (thick blue line) is $\propto t^{-0.3}$ as it is mostly powered by diffusion emission (thin blue line) [first term in equation (\ref{eq:Lbl KN})]. 
  After a transitional phase, later emission (thick red line) follows deposition rate of thermal energy through r-process (thin red line in the optically thin part; and dashed line in the whole ejecta) as $\propto t^{-1.3}$ [second term in equation (\ref{eq:Lbl KN})], and asymptotically $\propto t^{-2.3} - t^{-2.8}$ (for a similar figure see Figure 4 in \citealt{2019ApJ...876..128K}; also see  \citealt{2019ApJ...878...93W,2020ApJ...891..152H}).
  }
  \label{fig:LRP} 
\end{figure}

As shown in Figure \ref{fig:LRP}, at early times (typically $<1$ d) the diffusion part is largely dominant;
however, at later times, as the majority of the ejecta mass is exposed in the optically thin outer part, the second term takes over.

The time evolution of the term which represents the trapped thermal energy $E_{\rm{i}}(<\beta_{\rm{d}},t)$ at shells moving with velocities slower than $\beta_{\rm{d}}$ can be found by considering energy deposition by r-process and adiabatic cooling, giving ${\partial E_{\rm{i}}(<\beta_{\rm{d}},t)}/{\partial t}=-{E_{\rm{i}}(<\beta_{\rm{d}},t)}/{t}+f_{{\rm{tot}}}(t)\dot{E}_{\rm{dep}}(<\beta_{\rm{d}},t)$.
Hence:
\begin{equation}
    \begin{split}
    E_{\rm{i}}(<\beta_{\rm{d}},t) &=\frac{1}{t}\int_0^t f_{\rm{tot}}(t)\dot{E}_{\rm{tot}}(<\beta_{\rm{d}},t)tdt\\
    &\propto t^{1-k} f_{\rm{tot}}(t)M_{\rm{e}}(<\beta_{\rm{d}},t),
    \label{eq:M0.3}
    \end{split}
\end{equation}
where $\dot{E}_{\rm{dep}}(<\beta_{\rm{d}},t)=M_{\rm{e}}(<\beta_{\rm{d}},t)\dot{\varepsilon}_{\rm{0}}\left({t}/{\rm{1\:d}}\right)^{-k}$ is r-process energy deposition in all shells slower than $\beta_{\rm{d}}$ with a mass $M_{\rm{e}}(<\beta_{\rm{d}},t)$.

At early times, a key approximation is that $\beta_{\rm{d}}\sim \beta_{\rm{m}}$, and hence $\rho(\beta_{\rm{d}},t)\sim \rho(\beta_{\rm{m}},t)$ (see Section 3.1 in \citealt{2015ApJ...802..119K}).
From the condition of a sharp diffusion shell [equation (\ref{eq:tau})], $1/(\beta_{\rm{m}}-\beta_{\rm{d}})\propto \rho(\beta_{\rm{m}},t) t (\beta_{\rm{m}}-\beta_{\rm{d}})$, and the diffusion velocity moves inward through the ejecta so that $\beta_{\rm{m}} -\beta_{\rm{d}} \propto t$ [as $\rho(\beta_{\rm{m}},t)\propto t^{-3}$].
Consequently, during a time interval $\Delta t$, $(\beta_{\rm{m}} -\beta_{\rm{d}}) \propto \Delta t$, and the newly exposed mass $\Delta M_{\rm{e}}(<\beta_{\rm{d}},t) = \rho(\beta_{\rm{m}},t) \Delta V \propto \Delta t$.
Hence, $\Delta M_{\rm{e}}(<\beta_{\rm{d}},t)/{\Delta t} \sim \rm{Const}$.
In other words, at early times the diffusion shell moves constantly in the mass coordinate (see \citealt{2015ApJ...802..119K}; also see Appendix E in \citealt{2024ApJ...963..137H}).
Also, at early times $f_{\rm{tot}}(t)\equiv f_{\rm{tot}} \sim 0.7$ as density is high and most radioactive energy deposition is thermalized (except neutrinos).
Therefore, 
at early times the evolution of the kilonova luminosity $L_{\rm{KN}}(t)\sim L_{\rm{KN}}(<\beta_{\rm{d}},t) \sim {\Delta E_{\rm{i}}}/{\Delta t}$ can be found as [using equations (\ref{eq:Lbl KN}) and (\ref{eq:M0.3})]:
\begin{equation}
    L_{\rm{KN}}(t)\propto t^{1-k} \propto t^{-0.3} \:\:\:\:\:\: \text{[Early times]}.
    \label{eq:0.3}
\end{equation}
Hence, as shown in Figure \ref{fig:LRP}, at early times, the amount of trapped thermal energy to be released is expected to follow $\propto t^{-0.3}$.
It should be stressed that this time dependency is independent of parameters of the ejecta (such as $n$).
This is consistent with \cite{2015ApJ...802..119K} [see their equation (19) and Figure 3].

The time evolution of the second term simply follows the deposition rate of thermal energy through r-process (initially as $\propto t^{-k}$).
Early on, most radioactive energy is thermalized (except neutrinos which account for $\sim 30\%$ of the energy deposition; i.e., $f_{\rm{tot}}\sim 0.7$) [see \citealt{2014ApJ...789L..39W,2016ApJ...829..110B,2016MNRAS.459...35H,2017CQGra..34j4001R,2019ApJ...876..128K}; etc.].
At much later times as density decreases, $f_{\rm{tot}}$ drops as radioactive particles are less efficiently thermalized; hence the thermal energy deposition rate eventually enters its asymptotic phase and follows a steeper decline in the range $\propto t^{-2.3}$ (\citealt{2019ApJ...876..128K}) to $\propto t^{-2.8}$ (\citealt{2019ApJ...878...93W,2020ApJ...891..152H}).

\subsection{Application to GRB~211211A}
\label{sec:Application}
We aim to investigate the origin of the KN emission associated with GRB~211211A,
and whether it can entirely be explained by r-process powered KN emission (\citealt{2022Natur.612..223R,2022Natur.612..228T}).
We employ our analytical r-process KN model (see Section \ref{sec:KN model}; and Appendix E in \citealt{2024ApJ...963..137H} for a full description).
First, we search for a combination of two KN models (i.e., two ejecta components) capable of explaining the SED and the bolometric data: a blue KN model with low opacity (to explain the early blue emission) and a red KN model with high opacity (to explain the late red emission). 
We proceed as follows: first we carry out a parameter search to find models capable of explaining the early time data,
then we search for a complementary red KN model that explains the rest of data (late time data in particular). 

The main parameters for each KN model are $M_{\rm{e}}$ and $\kappa$ as the KN emission depends strongly on them.
11 values of $M_{\rm{e}}$ spread linearly in the interval $0.01\:M_\odot - 0.05\:M_\odot$,
and 11 values of $\kappa$ spread logarithmically in the interval $10^{-1} - 10$ cm$^{2}$ g$^{-1}$ (range of values is motivated by results of radiative transfer simulations with realistic atomic data in \citealt{2023arXiv230405810B}).
$\beta_{\rm{0}}$ is taken as 0.05 as suggested by post-merger mass ejection and GW170817 results (see \citealt{2018MNRAS.481.3423W,2018ApJ...860...64F}).
We take $\beta_{\rm{m}}=0.4$ (also $\beta_{\rm{m}} =0.3$, although the results are similar).
Finally, we take $n=2$ as expected from post-merger mass ejecta (i.e., constant mass ejection $\dot{M}_{\rm{e}}\propto \rm{Const.}$; $n=3.5$ has also been considered, but the results are similar).

We focus on two observational facts: 
i) the early ($5 - 10$ h) blue KN emission associated with GRB~211211A is quite luminous ($\sim 3-4\times 10^{42}$ erg s$^{-1}$; see Table \ref{tab:fit}; \citealt{2022Natur.612..228T});
ii) late observations at 4.4 d put an upper limit on the luminosity on the late red KN (in particular upper limit in R band by DOT; \citealt{2022Natur.612..228T}).
We search for combinations of KN models that can both reproduce the early brightness and that do not overpredict (overshoot) the late red KN emission.


\section{Results}
\label{sec:4}

\subsection{The light curve}
\label{sec:light curve}

To reproduce the bright bolometric luminosity of the early blue KN emission in GRB~211211A (see Table \ref{tab:fit} and \citealt{2022Natur.612..228T}), we search for viable r-process powered KN models.
Via an r-process powered KN model, the bright early emission can be achieved by increasing the ratio $M_{\rm{e}}/\kappa$;
as the photon diffusion time is $\propto\sqrt{\kappa M_{\rm{e}}}$, and r-process energy $E\propto M$, then $L\propto E/t \propto \sqrt{M_{\rm{e}}/\kappa}$.

We use $n=2$ which corresponds to a constant mass ejecta rate $\dot{M}_{\rm{e}}\propto \rm{Const}$.
As $L_{\rm{KN}}\propto t^{-0.3}$ at early times regardless of $n$, $n$ has a limited effect.
The impact of the parameter $\beta_{\rm{m}}$ is also limited.
This is because $\beta_{\rm{m}}$ has a physically limited range (as $<1$), and the temperature dependence on it in this range is not  strong.
Also, as the kinetic energy of the merger ejecta is $\propto  \beta_{\rm{m}}^2$, higher values ($\beta_{\rm{m}}\sim 0.8$) are not allowed as they imply very bright X-ray/radio emission from the merger ejecta, which has not been observed.
The impact of $\beta_{\rm{0}}$ is even weaker (as long as $\beta_{\rm{0}}\ll \beta_{\rm{m}}$).

Hence, either a low $\kappa$ or a high $M_{\rm{e}}$ parameter space is expected to explain the bright early KN emission.
However, it should be stressed that a KN model that overpredicts the late red KN emission (at $\sim 4$ d) should also be ruled out.
There are two important properties of r-process powered emission to recall: i) r-process energy deposition is $\propto t^{-1.3}$ and  early emission is dominated by diffusion of trapped thermal energy which has a shallow luminosity evolution [$\propto t^{-0.3}$ at early times; see equation (\ref{eq:0.3}); see Section \ref{sec:KN model} and Figure \ref{fig:LRP}]; and
ii) luminosity at late times scales to the ejecta mass ($\propto M_{\rm{e}}$). 
Hence, on one hand, employing a large ejecta mass model to explain the bright early KN emission can potentially overpredict the observed late red KN emission.
On the other hand, adopting a low opacity ($\kappa\sim 0.1$ cm$^{2}$ g$^{-1}$) and a less massive ejecta mass ($\sim 0.02\:M_\odot$) could avoid the issue of overpredicting the late red KN emission;
however, while this might reproduce the early bolometric luminosity, the emission would be shifted too far into bluer colors and will consequently underproduce (undershoot) the observed blue KN emission in optical bands. 
Adopting higher opacities ($\kappa\sim 1$ cm$^{2}$ g$^{-1}$) would avoid this problem but, again, at the expense of making the early bolometric luminosity [calculated using equation (E5) in Appendix E of \citealt{2024ApJ...963..137H}] too faint to explain the early observations (5 h $-$ 10 h epoch) [similar effects can be seen in Figure \ref{fig:FKN} for $M_{\rm{e}}\sim 0.04\:M_\odot$ and $\kappa \sim  0.1 - 10$ cm$^{2}$ g$^{-1}$].

\begin{figure}
    \centering
    \includegraphics[width=0.99\linewidth]{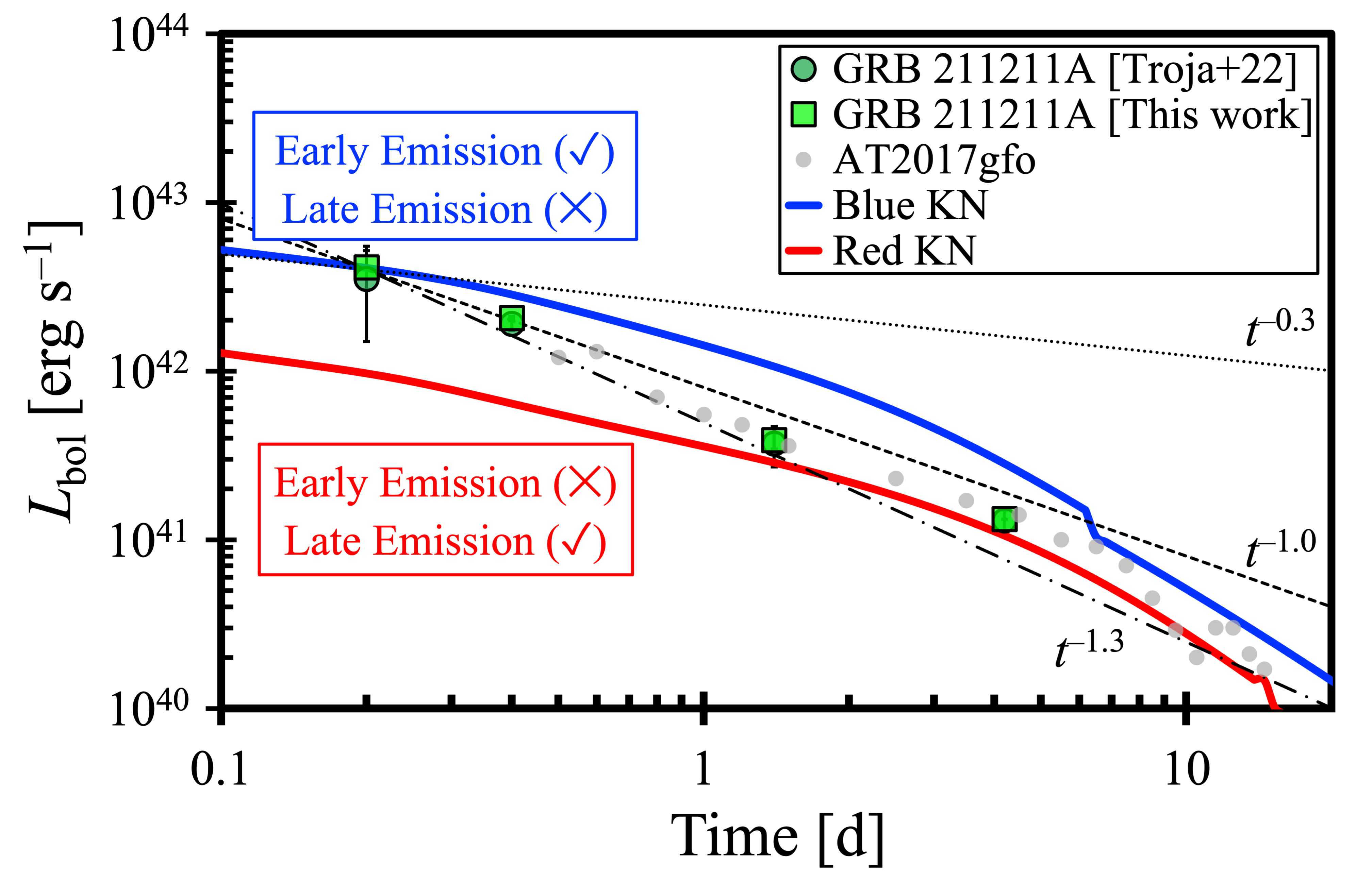} 
  \caption{The bolometric luminosity for a blue KN model (blue line; $\kappa = 1$ cm$^{2}$ g$^{-1}$ and $M_{\rm{e}}=0.05\: M_{\odot}$) that can explain the early KN emission in GRB~211211A (but overpredicts late KN emission); 
  and a red KN model (red line; $\kappa = 10$ cm$^{2}$ g$^{-1}$ and $M_{\rm{e}}=0.04\: M_{\odot}$) that can explain the late KN emission in GRB~211211A (but underpredicts late KN emission). 
  The observed KN emission in GRB~211211A (dark grey squares for our fit, and dark grey circles for \citealt{2022Natur.612..228T}; see Table \ref{tab:fit}) and AT2017gfo (grey circles; data from \citealt{2018MNRAS.481.3423W}) are shown.
  Dotted, dashed, and dotted dashed lines highlight time evolving power-law functions with indices of $-0.3$, $-1.0$, and $-1.3$ respectively. 
  This illustrates the difficulty of r-process powered KN models in explaining the fast time evolution of the KN emission in GRB~211211A.
  Other parameters of the KN models are $\beta_{\rm{0}}=0.05$, $\beta_{\rm{m}}=0.4$, and $n=2$.
  }
  \label{fig:LKN} 
\end{figure}

Figure \ref{fig:LKN} shows two light curves, the first from a blue KN model, and the second from a red KN model, chosen to reproduce early and late observations, respectively.
The tendency of the blue KN model to overpredict the observed late red emission, as well as the tendency of the late red KN model to underpredict the early blue emission, is apparent.
The overall time evolution of the observed KN emission (early to late KN emission in GRB~211211A; as well as AT2017gfo data) requiring a steeper power-law function (with an index between $-1$ and $-1.3$) than what is expected from r-process at early times [see equation (\ref{eq:0.3}) and Figure \ref{fig:LRP}] is also apparent.

\subsection{The SED}
\label{sec:SED}

\begin{figure*}
    \centering
    \includegraphics[width=0.55\linewidth]{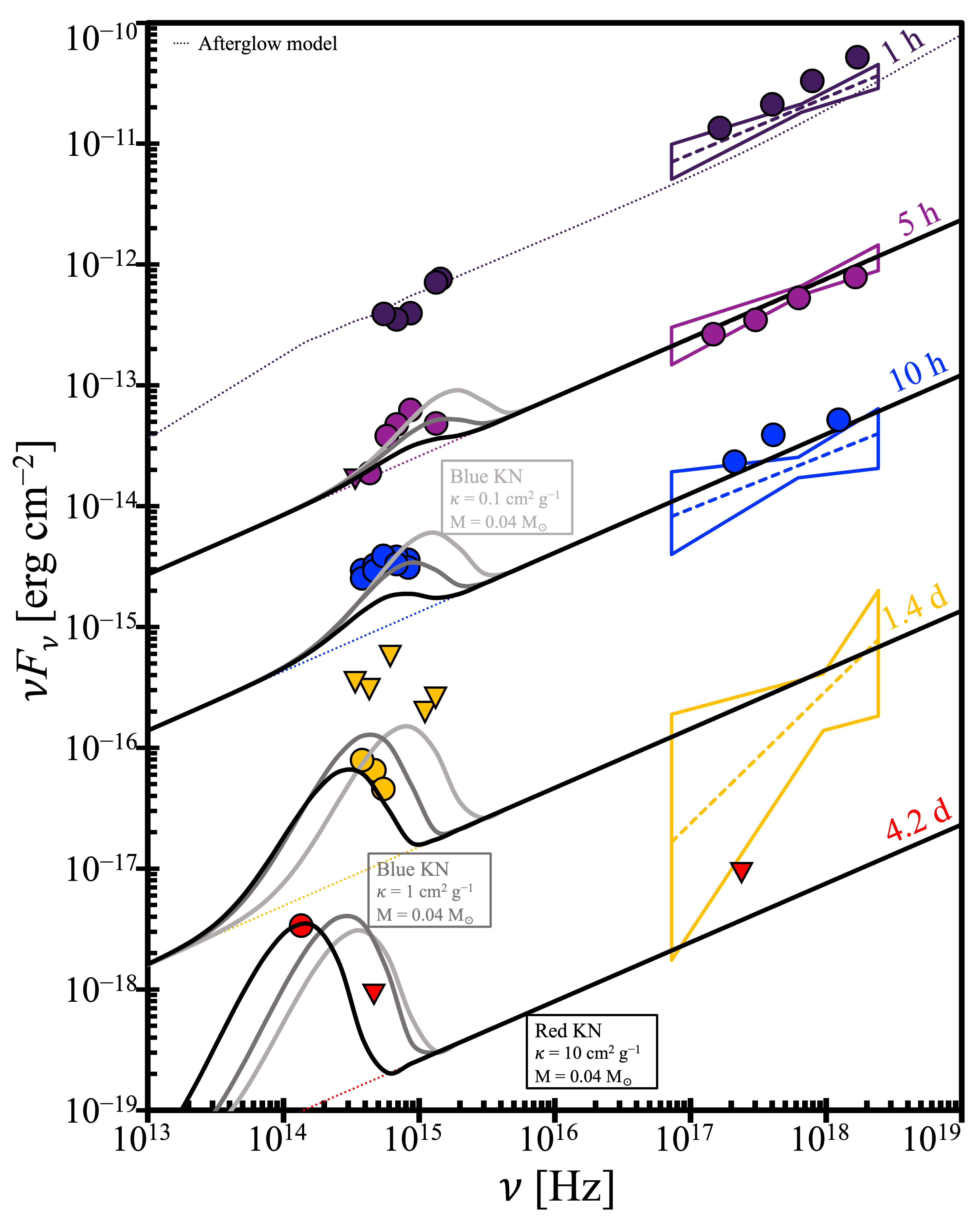} 
  \caption{SED of GRB~211211A and its time evolution.
  Five epochs are shown: 1 h (dark purple), 5 h (purple), 10 h (blue), 1.4 d (orange), and 4.2 d (red).
  Filled circles indicate detections, while triangles indicate upper limits. 
  Central values for the X-ray's photon index (dashed lines) and the corresponding incertitude (bow ties) are shown (see Appendix \ref{data}).
  The afterglow model is shown at each epoch with a dotted line.
  Three r-process KN models are shown (using $z =0.0762$): A low opacity blue KN model ($\kappa=0.1$ cm$^{2}$ g$^{-1}$ and $M_{\rm{e}}=0.04\: M_{\odot}$; light grey), a moderate opacity blue KN model ($\kappa=1$ cm$^{2}$ g$^{-1}$ and $M_{\rm{e}}=0.04\: M_{\odot}$; dark grey), and a red KN model ($\kappa=10$ cm$^{2}$ g$^{-1}$ and $M_{\rm{e}}=0.04\: M_{\odot}$; black).
  The inability of r-process powered blue KN models to explain both early observations without overproducing late emission is apparent.
  Also, the inconsistency of the low opacity model (light grey) with the color of early data is apparent.
  IR/Opt/UV data points were taken from \cite{2022Natur.612..223R,2022Natur.612..228T} and X-ray data was taken from \cite{2022Natur.612..228T}.
  For clarity data is scaled at each epoch by the following factors $10^0$, $10^{-0.8}$, $10^{-1.6}$, $10^{-2.4}$, and $10^{-3.2}$ (same as in \citealt{2022Natur.612..228T}).
  For a similar plot see Figure 2 in \cite{2022Natur.612..228T} and Extended Data Figure 2 in \cite{2022Natur.612..223R}.
  }
  \label{fig:FKN} 
\end{figure*}

Figure \ref{fig:FKN} shows the SED for an afterglow and the r-process powered KNe models with varying opacities ($\kappa=0.1$ cm$^{2}$ g$^{-1}$, $\kappa=1$ cm$^{2}$ g$^{-1}$, and $\kappa=10$ cm$^2$ g$^{-1}$, in light grey, dark grey, and black, respectively) and $M_{\rm{e}}=0.04\:M_\odot$.
As explained in Section \ref{sec:light curve}, low opacity tend to shift the SED to higher frequencies.
Consequently, an r-process powered blue KN model with low opacity ($\kappa=0.1$ cm$^{2}$ g$^{-1}$), although it can give a bright enough early blue emission, is inconsistent with early (and late) observations in terms of color (too blue).
A moderate opacity of $\kappa=1$ cm$^{2}$ g$^{-1}$, requires a higher mass to match early observations (5 h to 10 h);
however, such high mass is incompatible with the late observations (1.4~d to 4.2~d), in particular such models overshoot the R-band upper-limit at the 4.2 d epoch (more precisely at 4.4 d; see Table 1 in \citealt{2022Natur.612..228T}). 
Hence, we find that it is challenging to explain the entire data set even with a combination of two r-process powered KN models (blue KN and red KN models combined).

\subsection{Constraining $\kappa$ \& $M_{\rm{e}}$}
\label{sec:Constraints}
We investigate whether there are other possible r-process powered KN models, with different parameters, that could explain GRB~211211A's data set.
Our main parameters are $M_{\rm{e}}$ and $\kappa$; 
set to take 11 values each, in the range, $0.01 -0.05\: M_\odot$, $0.1-10$ cm$^{2}$ g$^{-1}$ respectively (see Section \ref{sec:Application}).
The other parameters ($\beta_{\rm{m}}$, $\beta_{\rm{0}}$, and $n$) have been explored individually but were found to have limited effects.

\begin{figure}
    \centering
    \includegraphics[width=0.99\linewidth]{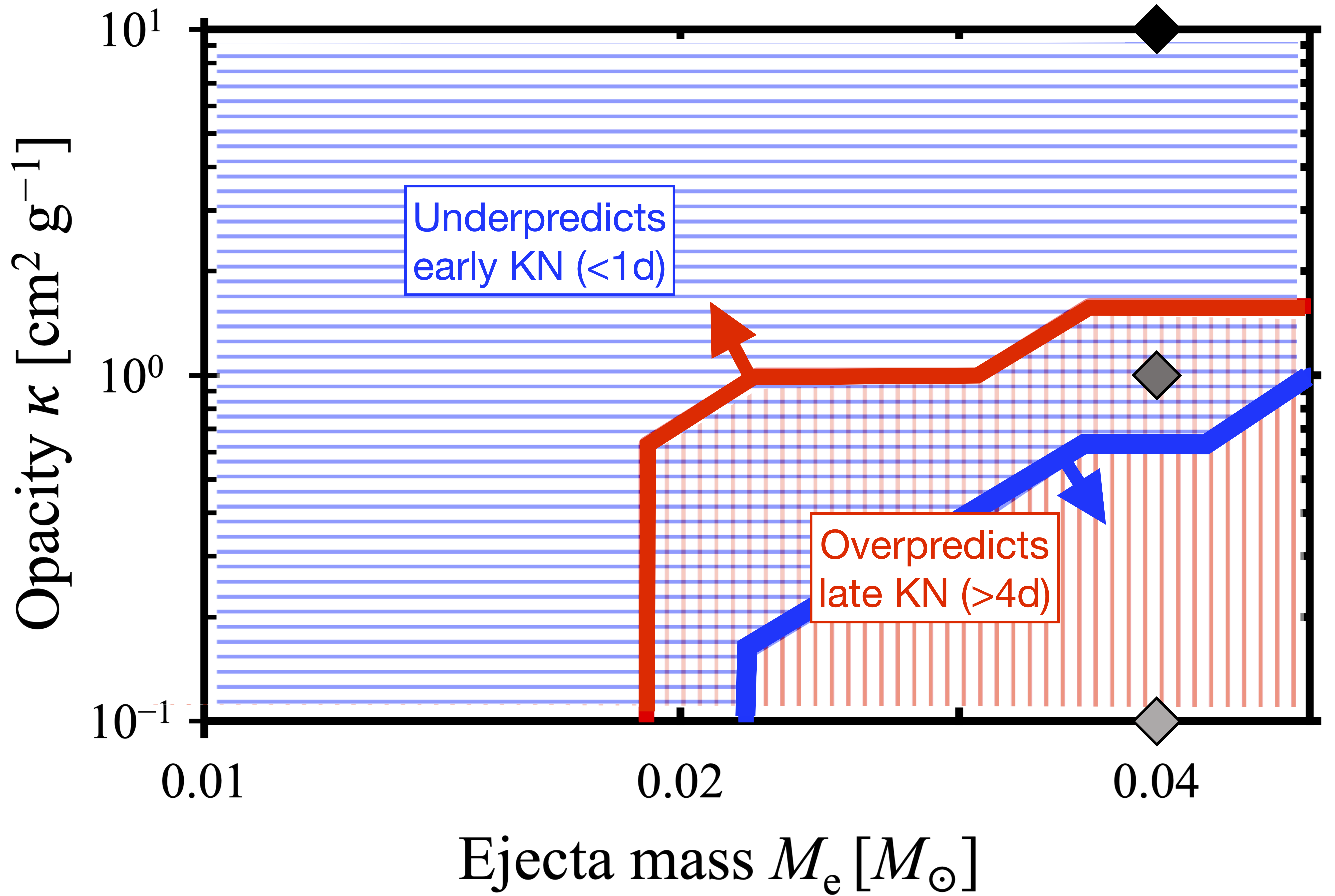} 
  \caption{The ejecta mass ($M_{\rm{e}}$) and opacity ($\kappa$) parameter space for r-process powered KN models. 
  Area constrained by the thick blue line (as indicated by the blue arrow) is where KN models are at least as bright as the early (5 h - 10 h) blue KN emission observed in GRB~211211A (within $1\sigma$); blue stripes indicate the parameter space where this is not the case (ruled out). 
  Area constrained by the thick red line (as indicated by the red arrow) is where the KN models do not overpredict the late time red emission (1.4 d $-$ 4.2 d; see Figure \ref{fig:FKN}); red stripes indicate the parameter space where this is not satisfied (ruled out). 
  There is no eligible parameter space where both the early and late emission of the KN in GRB~211211A can be explained by a combination of two r-process powered KN models.
  Diamond symbols indicate the three KN models shown in Figure \ref{fig:FKN}.
  The other parameters are taken as $\beta_{\rm{0}}=0.05$, $\beta_{\rm{m}}=0.4$ and $n=2$ (found to have limited effects on the results).
  }
  \label{fig:para} 
\end{figure}

Using our analytic model for r-process powered KN, and as a test of the hypothesis that the blue KN is r-process powered, 
we search for parameters where the KN associated with GRB~211211A can be explained in its entirety.
First, we evaluate the bolometric luminosity, and rule out the parameter space where the r-process powered KN luminosity at early times (5 h $-$ 10 h) is less than that of the KN in GRB~211211A (by more than $1\sigma$; see Table \ref{tab:fit}).
Then, we evaluate the observed $\nu F_{\nu}$, at late times ($\sim$1.4 d and $\sim$4.2 d), and rule out models that overshoot the data;
in particular, the $3\sigma$ R-band upper limit (by DOT; \citealt{2021GCN.31299....1G}) at 4.4 d poses a strict constraint (see Figure \ref{fig:FKN}; also see Figure 2 in \citealt{2022Natur.612..228T}).
It should be noted that these two criteria are quite conservative.

As shown in Figure \ref{fig:para}, we find no parameter space where both criteria are fulfilled.
In summary, the observed blue KN emission after GRB~211211A is so bright that in order to explain it via r-process, 
a large mass (and/or low $\kappa$) is required, which (due to the shallowness of the early KN light curve; see Figure \ref{fig:LRP}) ends up overpredicting (and contradicting) the late red KN emission data.
This result suggests a different origin for the early blue KN emission in GRB~211211A, other than ``r-process", such as the ``central engine" [see Section \ref{sec:4}].

\section{The Alternative: Central Engine powered KN}
\label{sec:5}
In Section \ref{sec:4}, we found that the r-process powered KN model has its limitations when explaining the early data in GRB~211211A;
here we explore an alternative scenario.

It is important to highlight two important observational facts.
First, observations of SGRBs (with a likely BNS merger origin) have consistently shown that, after the prompt phase, there is an extended/plateau phase (e.g., \citealt{2005ApJ...635L.133B,2006ApJ...643..266N}; \citealt{2013MNRAS.431.1745G}).
This late phase is present in the majority of SGRBs (\citealt{2017ApJ...846..142K}), and it has been associated with late engine activity (\citealt{2005ApJ...631..429I}).
Second, follow-up observations of GRB~211211A by Fermi-LAT detected GeV emission at $\sim 10^4$ s after the prompt emission (\citealt{2022Natur.612..236M}).  
This late GeV emission has been explained with late central engine activity, launching a late (long-lasting) low-power jet that interacts with the KN (\citealt{2019ApJ...887L..16K,2022Natur.612..236M}).
Hence, these two observational facts support late engine activity launching a low-power jet in BNS merger systems, such as GRB 211211A and other SGRBs.

In an attempt to investigate the origin of the early blue KN emission in GRB~211211A, and find an alternative to the r-process powered KN scenario, we consider late central engine activity. 
Considering the timescale of the GeV emission ($\sim 10^4$ s after), we adopt a plateau-like long-lasting ($\sim 10^4$ s) and low-power engine (i.e., $\sim 10\%$ radiative efficiency).
Since the typical luminosity of the plateau phase in X-rays is $\sim 10^{46}$ erg s$^{-1}$ (\citealt{2017ApJ...846..142K}), we consider a jet with a total power of  $ 10^{47}$ erg s$^{-1}$.
The late jet opening angle is not well understood from observations, however, considering that SGRB-jet opening angles (measured via jet break) are typically $\sim 6^{\circ}$ (\citealt{2023ApJ...959...13R}). 
We adopt a slightly larger, yet comparable, opening angle of $7.5^{\circ}$, as it has been suggested that the late jet takes a wider opening angle compared to the prompt jet \citep{2023MNRAS.522.5848L}.

In the central engine scenario, assuming the same ejecta as in the KN models, interaction of the central engine powered jet with the ejecta is considered.
This jet-ejecta interaction produces a shock that converts kinetic energy into thermal energy, in the form of a hot ``cocoon" component surrounding the jet.
As this thermal energy diffuses out of the ejecta, it produces the emission.
This emission is calculated analytically in two steps as follows.

In the first step, we solve the jet propagation through the merger ejecta (via jump conditions) using the analytic model in \cite{2020MNRAS.491.3192H,2021MNRAS.500..627H}, this allows us to estimate the time it takes the jet to break out of the ejecta, and estimate the amount of thermal energy produced via the jet-ejecta interaction in the form of a cocoon.
For this, we use the above jet parameters ($L_{\rm{iso,0}}=10^{47}$ erg s$^{-1}$ and $\theta_{\rm{0}}=7.5^\circ$) and the  ejecta parameters that correspond to the red KN model: $M_{\rm{e}}=0.04\:M_\odot$, $n=2$, $\beta_{\rm{m}}=0.4$, with the exception that we use a much smaller inner velocity $\beta_{\rm{0}} =\beta_{\rm{m}}/100$.
The smaller inner velocity $\beta_{\rm{0}}$ is motivated by the expected slower gravitationally bound component that is not relevant to the KN emission but is relevant to the jet propagation (at much earlier times).
Note that the mass of this slower component is negligible (as $n=2$ and $M\propto \beta$), and the density profile here is the same as that in the red KN model.

In the second step, diffusion emission from the thermal energy produced in the jet-ejecta interaction is calculated analytically following \cite{2023MNRAS.520.1111H,2023MNRAS.524.4841H}\footnote{It should be noted that, acceleration of ejecta shells due to energy supplied by the jet is assumed insignificant as the kinetic energy of the ejecta dominates (see \citealt{2023MNRAS.520.1111H}).}.
We focus on the diffusion emission from the cocoon trapped inside the ejecta ($<\beta_{\rm{m}}$) as this peaks at times relevant to the blue KN emission (see Appendix D in \citealt{2024ApJ...963..137H}).
In addition, we estimate the ram pressure balance between the shocked ejecta (trapped cocoon) and the unshocked ejecta, to determine the lateral spreading velocity ($\beta_{\perp}$) of the trapped cocoon;
in the case where $\beta_\perp \gtrsim \beta_{\rm{m}}$, the trapped cocoon is considered to spread and reach a spherical asymptotic geometry.
This was found to be the case for our central engine model\footnote{Here we did not consider the possibility of disintegration of r-process heavy element by the jet-cocoon shock \citep{2012ApJ...753...69H,2023arXiv230508575G} because this would be inconsistent with observations of the late red KN.}.

\begin{figure}
    \centering
    \includegraphics[width=0.99\linewidth]{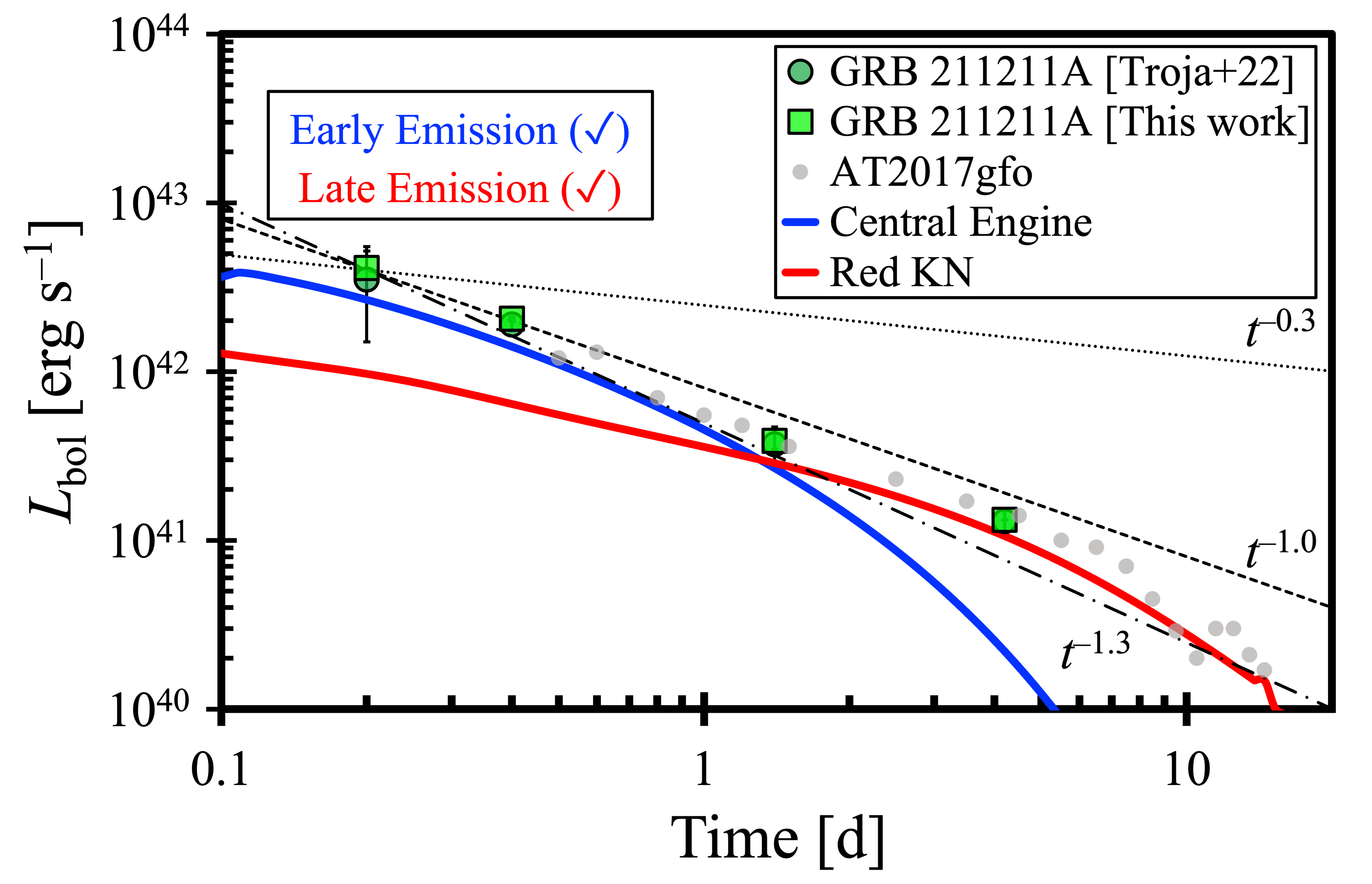} 
  \caption{Same as Figure \ref{fig:LKN}. The bolometric luminosity of our alternative central engine model where a low-power jet launched by late engine activity interacts with the lanthenide rich ejecta to explain the early ($<1$ d) KN emission in GRB~211211A (blue line), while r-process explains the late ($1-4$ d) red KN emission (red line) [see Figure \ref{fig:FCE} for the SED].
  This model can explain GRB~211211A data with only one red ejecta component.
  Parameters of the jet in the central engine model are: jet isotropic equivalent luminosity $L_{\rm{iso,0}}=10^{47}$ erg s$^{-1}$ and jet opening angle $\theta_{\rm{0}} = 7.5^\circ$.
  }
  \label{fig:LCE} 
\end{figure}

\begin{figure*}
    \centering
    \includegraphics[width=0.55\linewidth]{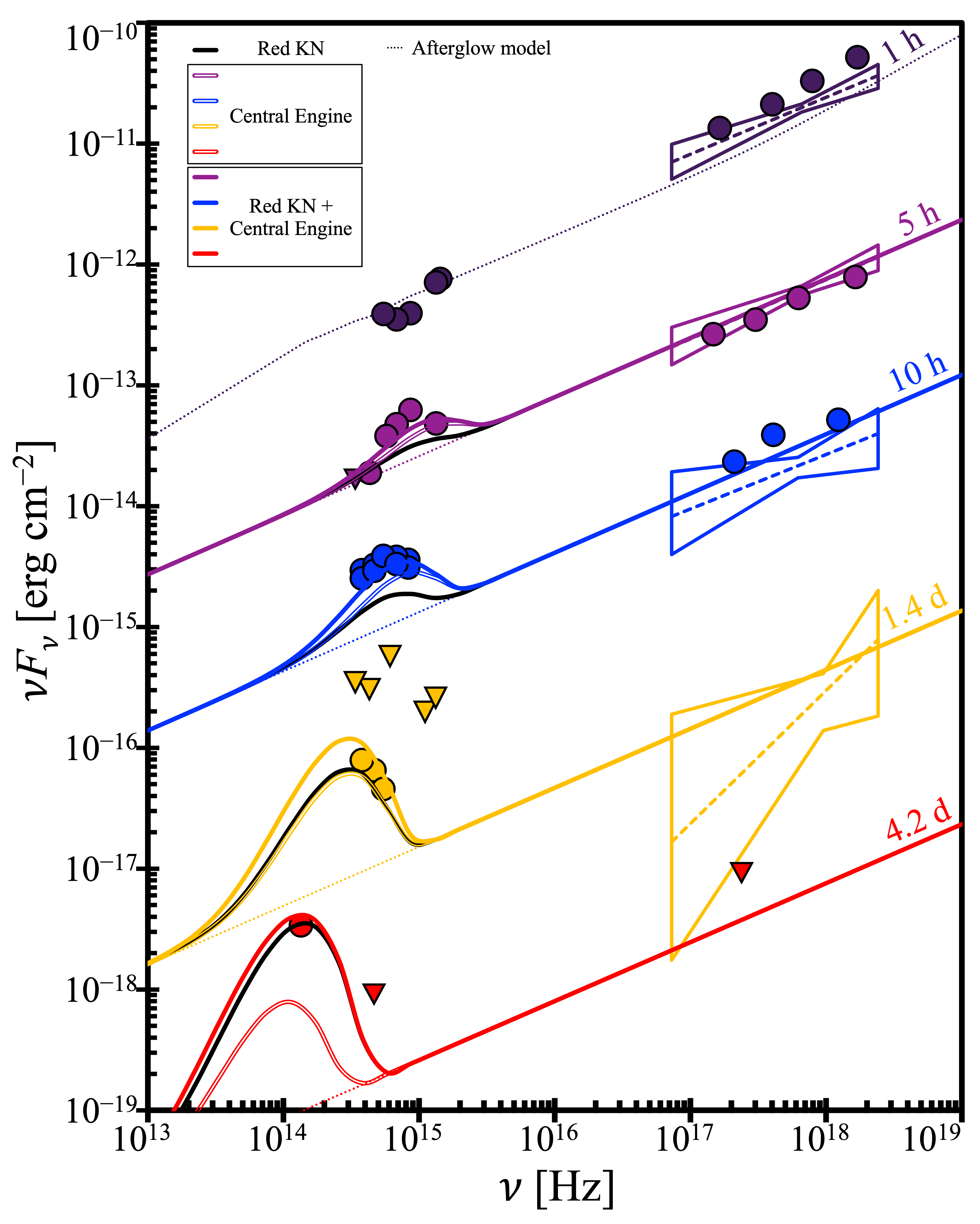} 
  \caption{Same as Figure \ref{fig:FKN}. 
  The SED of our alternative central engine model where a low-power jet launched by late engine activity interacts with the lanthenide rich ejecta to explain the early ($5-10$ h) blue KN emission in GRB~211211A (double lines), while r-process explains the late ($1.4-4.2$ d) red KN emission (black line; same model as in Figures \ref{fig:LKN} and \ref{fig:FKN}). 
  The central engine model, combined with the red KN model (solid colored lines), can explain both early and late emission data [see Figure \ref{fig:FCE} for the bolometric luminosity].
  }
  \label{fig:FCE} 
\end{figure*}

Diffusion emission from the thermal energy deposited by the jet-ejecta interaction gives an additional luminosity term ($L_{\rm{CE}}$) in addition to the two r-process terms in equation (\ref{eq:Lbl KN}) so that:
\begin{equation}
    L_{{\rm{tot}}}(t)=L_{{\rm{KN}}}(<\beta_{\rm{d}},t)+L_{{\rm{KN}}}(\geqslant\beta_{\rm{d}},t) + L_{{\rm{CE}}}(<\beta_{\rm{d}},t)  ,
        \label{eq:Lbl KN+CE}
\end{equation}
and $L_{{\rm{CE}}}$ can be found as [see equation (12) in \citealt{2024ApJ...963..137H}; for more details see their Appendix D]:
\begin{equation}
\begin{split}
&L_{{\rm{CE}}}\approx 2.6\times10^{42} \text{ erg s$^{-1}$}\\
&\left(\frac{\theta_{\rm{0}}}{7.5^{\circ}}\right)^{2} 
\left(\frac{L_{{\rm{iso,0}}}}{10^{47}\text{ erg s$^{-1}$}}\right)
\left(\frac{t_{\rm{b}}}{4.7\times 10^3 \text{ s}}\right)\left(\frac{\kappa}{\text{$10$ cm$^{2}$ g$^{-1}$}}\right)^{\frac{p-2}{2}}\\
&\left(\frac{M_{\rm{e}}}{0.04\msun}\right)^{\frac{p-2}{2}}
\left(\frac{t_{\rm{obs}}}{5 \text{ h}}\right)^{-p}, \\
\end{split}
\label{eq:Lbl PL-T}
\end{equation}
where the index $p$ take values as $p\sim 1 $ (at $\sim 5$ h $- 10$ h) to $p\sim 2$ (at $\sim 1$ d $- 4.2$ d) which results in a steeper/faster time evolution than in the r-process model at early times [see equation (\ref{eq:0.3})].
We use the exact same ejecta parameters as those of the red KN model (shown in Figures \ref{fig:LKN}, \ref{fig:FKN}, \ref{fig:LCE}, and \ref{fig:FCE}) to calculate $L_{{\rm{CE}}}$: $\kappa=10$ cm$^2$ g$^{-1}$, $\beta_{\rm{m}}=0.4$, $n=2$, and $\beta_{\rm{0}}=0.05$.
Noting that this choice of parameters is for convenience, and should not be regarded as unique for central engine model.

In summary, with this central engine scenario, no additional KN components are required and
only a single red ejecta component is used. 
Here the new consideration is the additional thermal energy from the jet-ejecta interaction.

Figure \ref{fig:LCE} shows the bolometric light curve of our central engine powered model.
In comparison to the r-process powered blue KN model (see Figure \ref{fig:LKN}), the central engine model light curve has a steeper decay.
This fast evolution at early times is in contrast with the r-process model,
as the issue of overpredicting the late red KN emission is avoided (see Section \ref{sec:light curve}).
This major difference is due to the thermal energy deposition at much earlier times relative to diffusion timescale in the central engine model; 
whereas thermal energy is constantly being supplied to the system in the r-process model ($\propto t^{-1.3}$; see Section \ref{sec:KN model}).

Although the late low-power jet has in total an energy budget that is on the order of $\sim \text{a few }\%$\footnote{This can be found considering that the the isotropic equivalent energy of the prompt emission of GRB~211211A is $\sim 5\times 10^{51}$ erg s$^{-1}$ (\citealt{2022Natur.612..232Y}), and that typical luminosity and duration of the average plateau phase gives an energy of $\sim 10^{46}\times 10^4\sim 10^{50}$ erg (\citealt{2017ApJ...846..142K}). Radiation efficiencies were assumed to be comparable ($\eta_\gamma\sim 10\%$; \citealt{2020MNRAS.493..783M,2022Natur.612..236M,2023ApJ...959...13R,2024ApJ...963..137H}). } of that of the prompt jet, its blue KN-like emission is bright (e.g., compared to the prompt jet's cocoon; see \citealt{2024ApJ...963..137H}; also see \citealt{2017ApJ...834...28N,2018MNRAS.473..576G}; and the ``shock-cooling" model in \citealt{2018ApJ...855..103P,2018ApJ...855L..23A}; for the prompt jet case).
This is because: i) the system is expanding homologously and thermal energy is subjected to adiabatic cooling (as $\propto V^{-\frac{1}{3}}\propto t^{-1}$);
and ii) the blue KN peaks at $\sim 1$ d $\sim 10^5$ s.
Therefore, by the peak time, ejecta heated by the prompt jet (launched at $ \sim 1$ s) would have cooled down adiabatically by a factor of $\sim 10^{-5}$; 
whereas, ejecta heated by the late jet (launched at $\sim 10^4$ s) cools down only by a factor of $\sim 10^{-1}$, which combined with the low energy budget of the late jet would still make the late jet model $\sim 10-100$ times brighter than the prompt jet model.

In Figure \ref{fig:FCE} the SED of the central engine powered model is presented (double lines).
The fast temporal evolution of the central engine model is noticeable (to be contrasted with r-process models in Figure \ref{fig:FKN}).
Combined with the r-process powered red KN model, the central engine model is consistent with almost all data points at all epochs.

Besides naturally explaining the KN associated with GRB~211211A at all epochs, our central engine model is appealing in other aspects.
Firstly, it is consistent with detection of extended/plateau emission in SGRBs;
this is also coherent with the detection of GeV emission in GRB~211211A and associated with the same type of jet (\citealt{2022Natur.612..236M}).
Hence, the consideration of a late jet is quite reasonable.
In addition, the energy requirement for the late jet is just $\sim \text{a few }\%$ of that of the prompt jet. 
This could naturally be explained by late time accretion onto the merger remnant (e.g., \citealt{2015ApJ...804L..16K,2023ApJ...958L..33G}; also \citealt{2023MNRAS.522.5848L} showed that this late jet is wider than the prompt jet).

Secondly, it only requires one single ejecta component.
Here, this single ejecta component is red (i.e., lanthenide rich; \citealt{2018MNRAS.481.3423W}), 
which is consistent with numerical relatively calculations showing that the red dynamical ejecta is faster, shielding the bluer post-merger ejecta, and giving the impression of an effectively red ejecta (\citealt{2018ApJ...865L..21K}). 
There is no need to invoke a second blue (or even a third purple) component.
In fact, they would typically require parameters that are not trivial considering numerical relativity calculation results (see \citealt{2017PhRvD..96l3012S,2018ApJ...860...64F,2019EPJA...55..203S});
although recently other physical mechanisms have been suggested (e.g., \citealt{2019PhRvD.100b3008M,2021PhRvD.104f3026S,2023ApJ...951L..12J}; also see Figure 9 in \citealt{2023MNRAS.525.3384K}).

The scenario of GRB central engine powered KN has been proposed in \cite{2015ApJ...802..119K,2016ApJ...818..104K} in general terms (and in \citealt{2018PTEP.2018d3E02I,2018ApJ...861...55M} in the context of AT2017gfo/GW170817).
It has been suggested that the prompt jet can affect the color of the KN emission (bluer \citealt{2020ApJ...900L..35C,2021MNRAS.500.1772N,2023PhRvL.131w1402C}; or redder \citealt{2023MNRAS.523.2990S}).
\cite{2022Natur.612..228T} pointed out the possibility of the GRB-jet's contribution to the KN associated with GRB~211211A.
\cite{2024ApJ...963..112M} argued for the same from statistical fitting of \cite{2022Natur.612..228T}'s results using a one-zone cocoon model.
However, their model is not reasonable as it assumes that the prompt jet somehow shocks most of the ejecta ($\sim 0.01\:M_\odot$) in the timescale of $\sim 1$ s (inconsistent with numerical simulations: \citealt{2014ApJ...788L...8M,2014ApJ...784L..28N,2018MNRAS.473..576G,2021MNRAS.500..627H}; etc.) depositing large amounts of thermal energy (also inconsistent with numerical simulations showing that kinetic energy dominates; see Figure 1 in \citealt{2021MNRAS.500..627H}), 
which gives inappropriate luminosity and temperature estimates when compared with numerical simulation estimates (\citealt{2018MNRAS.473..576G,2023ApJ...953L..11G}) and realistic analytic modeling (in particular; see Section 4.3.2 in \citealt{2023MNRAS.520.1111H}; Section 5.5 \citealt{2023MNRAS.524.4841H}).
Here, we demonstrate that energy injection from a long-lived central engine can naturally explain the KN associated with GRB~211211A;
with the crucial difference being, instead of the prompt jet (\citealt{2017ApJ...834...28N,2018ApJ...855..103P,2018ApJ...855L..23A}; etc.), we suggest a late low-power plateau-like jet with a reasonable energy budget of $\sim \text{a few }\%$ of the prompt jet's energy.


\section{Discussion \& Conclusion}
\label{sec:6}
Here, we revisited the early blue KN emission in GRB~211211A to better investigate its origin, having been previously explained using r-process heating.

We have presented a fully analytic KN model (Section \ref{sec:KN model}) and explained that r-process powered KNe follow a shallow temporal evolution at early times ($L_{{\rm{KN}}}\propto t^{-0.3}$) due to the continuous r-process energy deposition and the dominance of the diffusion of trapped thermal energy at early times (see Figure \ref{fig:LRP}).
We then applied our model to the KN emission associated with GRB~211211A (Section \ref{sec:Application}).

Our results indicate that the light curve of the early blue KN emission in GRB~211211A has a temporal evolution that is too fast to be explained via our analytic models (Section \ref{sec:light curve}), in the parameter space: $M_{\rm{e}}\sim 0.01\:M_\odot -0.05\:M_\odot$ (KN models with $M_{\rm{e}} < 0.01\:M_\odot$ or $M_{\rm{e}} > 0.05\:M_\odot$ are already excluded), and $\kappa \sim 0.1 - 10$ cm$^{2}$ g$^{-1}$, $\beta_{\rm{m}}\sim 0.4-0.5$, and $n\sim 2 -3.5$.
The main issue being that the early data is too bright, and in order to explain this via the r-process powered blue KN models, the required mass is large so that it leads to an overprediction of the red KN emission observed at late times;
whereas the red KN model is too dim at early times to explain the early bright blue KN emission (see Figure \ref{fig:LKN}). 
Employing low-opacity low-mass blue KN models is not ideal either, as the resulting colors are too blue to be consistent with early data (see Figure \ref{fig:FKN}).
As a result, over our wide parameter space, we did not find any combination of two r-process powered KN models (i.e., two component) that could explain the entire data set of GRB~211211A (see Figure \ref{fig:para}). 

We argue that the early data (5 $-$ 10 h) in GRB~211211A, may not be predominantly r-process powered.
Our alternative is a central engine powered KN.
We suggest that a low-power jet from late engine activity that interacts with the merger ejecta, offers a more natural explanation;
which is consistent with observations indicating that the majority of SGRBs have late long-lasting extended/plateau emission phases after the prompt emission (\citealt{2017ApJ...846..142K}).
Also, GeV emission observed $\sim 10^4$ s after the GRB~211211A has been explained with the same type of engine activity, jet, and timescale (\citealt{2022Natur.612..236M}).
We showed that such a low-power jet ($\sim 10^{46}$ erg s$^{-1}$ in X-ray) with its typical opening angle ($\sim 7.5^\circ$), explains naturally the bolometric light curve (see Figure \ref{fig:LCE}), and the SED at all epochs (see Figure \ref{fig:FCE}).

Hence, our conclusion is that the early blue KN emission in GRB~211211A hints to late engine activity.
This suggests that after the prompt emission, a late low-power engine activity phase may be quite common in SGRBs, and also potentially in LGRBs (as X-ray flares; \citealt{Nousek2006ApJ...642..389N}).
We argue that the interaction of this late jet with the surrounding ejecta could not have been identified in most standard/cosmological GRBs due to its faintness at large distances.
In the multi-messenger event GW170817/AT2017gfo, the first observations started at $\sim 10$ h, hence any early blue emission has been missed.
However, GRB~211211A as a nearby, well observed event, may have opened a new window to indirectly probe the evolution of the central engine of GRBs at later times after the prompt phase. 
\cite{2020MNRAS.493.3379R} suggested that several KNe candidates associate with SGRBs show exceptionally bright blue KNe (e.g., GRB~050724, GRB~060614, and GRB~070714B) while their red KNe are typical;
this is challenging to explain with r-process heating but can be explained naturally with our scenario of a late low-power jet.

Our result shows that a typical late-low-power jet model ($\sim 10^{46}$ erg s$^{-1}$ in X-ray; and $\sim 7.5^{\circ}$) synergies well with the red ejecta component to explain the blue KN.
However, it should be noted that alternative jet models may be viable.
Different central engine powered models such as the magnetar model have not been investigated here (\citealt{2018ApJ...856..101M,2013ApJ...776L..40Y}).
However, high neutrino radiation from the deferentially rotating hyper massive NS remnant can be a potential issue, as neutrinos increase the electron fraction ($Y_{\rm{e}}$) of the ejecta and suppress nucleosynthesis of heavy elements which would be at odds with the red KN emission (\citealt{2014MNRAS.441.3444M})\footnote{It should be noted that recent works indicate that even in cases where the remnant is a BH the disk outflow tend to have a higher $Y_{\rm{e}}$ (see  \citealt{2020PhRvD.101h3029F,2022MNRAS.509.1377J}) and hence, the formation mechanism of high-opacity massive ejecta is still being investigated; although MHD effects may favor lower $Y_{\rm{e}}$ values (\citealt{2023PhRvL.131a1401K}).}.
Additionally, the magnetar model could face the issue of producing a KN that is too bright when compared to the observations (\citealt{2024MNRAS.527.5166W}), and even if the magnetar outflow is collimated and on-axis (\citealt{2024MNRAS.528.3705W}), GRB~211211A would be expected to have a much brighter KN.
Finally, the magnetar model may not have as coherent an explanation for the GeV emission in GRB~211211A as our low-power-jet model (\citealt{2022Natur.612..236M}).

Our main finding here is that the current r-process powered blue KN models struggle at explaining the observed emission after GRB~211211A.
Here, we used a simplified model of the heating rate ($\propto t^{-1.3}$), however, it should be stressed that r-process models are still incomplete and uncertain (\citealt{2021ApJ...906...94Z,2021ApJ...918...44B,2024arXiv240403699M})\footnote{Is it worth noting that with revised heating rates, \cite{2024arXiv240407271S} found that the decline rates of KN  are typically overestimated (i.e., they decline more slowly than classically considered with simple models). This further supports our conclusion.}.
For instance, the heating rate could have a steeper decay for some very specific models (with high $Y_{\rm{e}}$, see Figure 5 in \citealt{2014ApJ...789L..39W}). 
Although, such models are not typical, they could present an alternative explanation for the KN emission following GRB~211211A.
It should be stressed that we do not rule out the existence of the r-process powered blue KN, or the existence of a blue (or purple) ejecta component.
Instead, we indicate that it is subdominant in luminosity and cannot fully explain the observed early emission. 
Hence, it could still coexist with the engine-powered emission in our scenario.

It should also be noted, that in terms of radiative transfer, our result relied on two simplifications: grey opacity (although the adopted values are compatible with realistic radioactive transfer simulations \citealt{2023arXiv230405810B}), and a single-temperature blackbody model.
In reality, radiative transfer in the KN ejecta can be more complex, which could result in a more fluctuating light curve (although at short timescales; see \citealt{2023arXiv230405810B}), and reprocessing of radiation.

Also, we explored a wide parameter space (in particular for $\kappa$ and $M_{\rm{e}}$) that, with the current understanding, is expected to cover a typical BNS event; 
however, considering the diversity in BNS and BH-NS mergers, extreme parameters that we did not cover could be possible in nature (e.g., see \citealt{2024arXiv240415027K}).

Finally, despite the KN of GRB~211211A being nearby and well observed, there are gaps in observations (in particular the SED), and systematic uncertainties; especially considering uncertainties in afterglow modeling (\citealt{2022Univ....8..612L}).
Hence, although we disfavor the current r-process model for the current data set of GRB~211211A, more research (e.g., nuclear physics and r-process nucleosynthesis) and more observations are needed to reach a more general conclusion on SGRBs. 

With more GRB-related missions available, such as Einstein Probe (\citealt{2015arXiv150607735Y}) and SVOM (\citealt{2015arXiv151203323C}), and in the near future ULTRASAT (\citealt{2024ApJ...964...74S}), HiZ-GUNDAM (\citealt{2020SPIE11444E..2ZY}), THESEUS (\citealt{2018AdSpR..62..191A}), etc., the prospect of more, and very early observations/follow-ups of GRBs, BNS/BH-NS mergers, and X-ray transients is promising.
Our proposed scenario of engine powered early kilonova, can be tested  with such future observations; together with the scenario of r-process powered KN.
Also, our scenario of late engine activity is very relevant to neutrino emission and could be tested with future neutrino observations \cite{2024arXiv240507695M}.


\section{Data availability}
The data underlying this article will be shared on reasonable request to the corresponding author.

\begin{acknowledgements}
We thank
    Alessio Mei,
    Ayari Kitamura,
    Clément Pellouin,
    Kazumi Kashiyama,  
    Kunihito Ioka,
    Nanae Domoto,
    Norita Kawanaka,
    Om S. Salafia,
    Sho Fujibayashi,
    Smaranika Banerjee,
    Tomoki Wada,
    and Wataru Ishizaki
    for their fruitful discussions and comments. 

    This research was supported by Japan Science and Technology Agency (JST) FOREST Program (Grant Number JPMJFR212Y),
    the Japan Society for the Promotion of Science (JSPS) Grant-in-Aid for Scientific Research (19H00694, 20H00158, 20H00179, 21H04997, 23H00127, 23H04894, 23H04891, 23H05432, and 23K19059), 
    JSPS Bilateral Joint Research Project, and National Institute for Fusion Science (NIFS) Collaborative Research Program (NIFS22KIIF005).
  
    This work was partly supported by 
    JSPS KAKENHI nos. 22K14028, 21H04487, and 23H04899 (S.S.K.). 
    S.S.K. acknowledges the support by  the Tohoku Initiative for Fostering Global Researchers for Interdisciplinary Sciences (TI-FRIS) of MEXT’s Strategic Professional Development Program for Young Researchers.
    G.P.L. is supported by the Royal Society via a Dorothy Hodgkin Fellowship (grant Nos. DHF-R1-221175 and DHF-ERE-221005).

    Numerical computations were achieved thanks to the following: Cray XC50 of the Center for Computational Astrophysics at the National Astronomical Observatory of Japan, and Cray XC40 at the Yukawa Institute Computer Facility.
\end{acknowledgements}

\bibliography{0-P7}
\bibliographystyle{aasjournal}


\appendix
\section{Data of the KN in GRB~211211A}
\label{data}
GRB~211211A's photometric data  infrared/optical/UV data is gathered from the literature (\citealt{2022Natur.612..223R,2022Natur.612..228T}).
Similarly to \cite{2022Natur.612..228T} we consider 5 epochs as a function of the observed time (since the GRB): 1 h, 5 h, 10 h, 1.4 d, and 4.2 d.

X-ray data is obtained from Swift's burst analyser webpage\footnote{\url{https://www.swift.ac.uk/burst_analyser/01088940/}}.
X-ray data is integrated over the intervals  $3.5\times10^3$ s $-$ $5\times10^3$ s, $15.9\times10^3$ s $-$ $22.1\times10^3$, $20.0\times10^3$ $-$ $64.8\times10^3$, and $50.0\times10^3$ s $-$ $300.0\times10^3$ s, so that the logarithmic central times correspond to the times of each respective epoch (all epochs except 4.2 d).
For each epoch we determine the photon index ($\Gamma$) and the normalization level.
Considering uncertainties in both parameters, X-ray observations are represented in the form of bow ties (see Figures \ref{fig:FKN} and \ref{fig:FCE}).

In addition, X-ray data points in \cite{2022Natur.612..228T} are used as a reference; except for the 1.4 d epoch (where the photon arrival time is not well consistent with the epoch time) all epochs are considered.

\subsection{Afterglow model}
\label{sec:AG}

\cite{2022Natur.612..223R}, using a Markov Chain Monte Carlo, fit a decelerating, relativistic forward shock with synchrotron emission model (see, \citealt{2018MNRAS.481.2581L}) to the afterglow dominated data from GRB 211211A. 
The fit model parameters are determined via the X-ray and radio afterglow data, with the early optical/near-infrared providing a spectral energy distribution constraint to the model and the later ($>0.1$ d) KN dominated optical/NIR data providing upper limits for the afterglow model. 
This light curve fit allows for the subtraction of the afterglow contribution to the kilonova dominated optical/NIR data. 
The optical afterglow emission at $>0.1$ d is within the self-similar deceleration phase and behaves as a simple power-law decline governed by the spectral and temporal indices inferred from the X-ray to radio broadband data.

The afterglow model assumes a uniform ``top-hat" jet structure viewed on-axis -- where the bright GRB 211211A supports the “on-axis viewer” assumption. 
The observable afterglow emission from a jet with an ultra-relativistic velocity, where the Lorentz factor is $\sim70$ (\citealt{2022Natur.612..223R}), is emitted from within a narrow cone with an angle defined by the inverse of the Lorentz factor, initially $<1^\circ$. 
Any wide-angle jet structure will be hidden to an on-axis observer until the jet break time when, as the jet decelerates, the inverse of the Lorentz factor becomes comparable to the jet core half opening angle. 
At the jet break time, the afterglow decline will typically steepen, and the shape of the jet break (how sharp/rapid the jet break is) contains information about the extent of any wider angled jet structure (\citealt{2021MNRAS.506.4163L}). 
Observations rarely have sufficient cadence and sensitivity at the jet break time to accurately determine the sharpness of the light curve change, thus the top-hat jet structure assumption is a valid approximation for most bright, and therefore on-axis viewed, GRBs.

The SED of our afterglow model have been compared to the X-ray data.
Apart from epoch 4 (1.4 d) where X-ray became too faint, our fit is consistent with the measured X-ray photon index.
Our afterglow model is also consistent with IR/Opt/UV data (see dotted lines in Figure \ref{fig:FKN}).

Our afterglow modeling confirms the presence of the thermal excess previously reported by \cite{2022Natur.612..223R,2022Natur.612..228T} [see Appendix \ref{sec:excess}]. 
We assessed the effect of uncertainties in the afterglow modeling on the flux of the thermal excess and confirmed that, within $1\sigma$ errors, their impact on the thermal excess is negligible (less than 10$\%$). 
Thus, this afterglow model, with the previously mentioned reasonable assumptions, robustly predicts the presence of a thermal component.

\subsection{Bolometric luminosity}
\label{sec:excess}
After subtracting the afterglow contribution, excess in IR/Opt/UV that is interpreted as KN emission is found.
Assuming a blackbody emission model, we determine the best fit (temperature and photospheric radius) for the excess.
Our results are shown in Table \ref{tab:fit} in comparison to those of \cite{2022Natur.612..228T}.
Our fit is very consistent with \cite{2022Natur.612..228T}, despite differences in afterglow modeling.
It should be noted that, due to the limited spectral coverage (in particular in UV at later epochs), this bolometric luminosity should be considered as a lower limit for the KN emission in the form of afterglow excess.

It should also be noted that, as our fit is statistical, fitting results (photospheric radius in particular) do not necessarily have a robust physical meaning.
For instance, in our physical model (that could explain the data) the maximum physical photospheric velocity $\beta_{{\rm{ph}}}$ is $<0.4$, while the fitting results indicates photospheric velocities $\beta_{{\rm{ph}}}>0.4$ at early times.
Hence, these photospheric values should not be taken at face value for their physical meaning.

\begin{table*}
\centering
\caption{
Parameters of the best-fit for the afterglow excess in GRB~211211A, in comparison to fitting results in \cite{2022Natur.612..228T}. 
1-$\sigma$ errors are given. 
Errors in the 4.2 d epoch could not be statistically evaluated due to the limited data (only two data points).
}
\label{tab:fit}
\begin{tabular}{c|ccc|ccc} 
\hline
\hline
& \multicolumn{3}{|c}{Our fit} & \multicolumn{3}{|c}{Troja+22}\\ 
\hline
Time & $L_{\rm{bol}}$ & $T_{\rm{}}$  & $r_{\rm{ph}}$ &  $L_{\rm{bol}}$& $T_{\rm{}}$  & $r_{\rm{ph}}$ \\
\text{[d]} &  [$10^{42}$ erg s$^{-1}$] & [$10^{3}$ K] &  [$10^{15}$ cm] &  [$10^{42}$ erg s$^{-1}$] & [$10^{3}$ K] &  [$10^{15}$ cm]\\
\hline
0.2 & $4.11 \pm 1.07$ & $14.55 \pm 3.32$ & $0.36 \pm 0.13$& $3.50 \pm 2.00$&$16.00 \pm 5.00$&$0.28 \pm 0.14$\\
0.4 & $2.05  \pm 0.06$ & $8.04 \pm  0.28$ & $0.83 \pm 0.056$& $1.90 \pm 0.15$&$8.00 \pm  0.30$&$0.80  \pm 0.05$\\
1.4 &  $0.39  \pm 0.04$ & $3.98 \pm  0.18$ & $1.47 \pm  0.21$& $0.37 \pm 0.10$&$4.90 \pm   0.50$&$0.90  \pm  0.20$ \\
4.2 &  $0.13$ & $2.68$ & $1.90 $& $0.13 \pm 0.02$&$2.50 \pm 0.10$& $2.00 \pm 0.20$\\
\hline
\end{tabular}
\end{table*}

\label{lastpage}

\listofchanges

\end{document}